# Comment to "Nonreciprocal Light Propagation in a Silicon Photonic Circuit"

Roel Baets[1,2*], Wolfgang Freude[3,4], Andrea Melloni[5], Miloš Popović[6], Mathias Vanwolleghem[7,8]

[1]Photonics Research Group, INTEC-department, Ghent University-IMEC, B-9000 Gent, Belgium
[2]Center for Nano- and Biophotonics, Ghent University, B-9000 Gent, Belgium

[3]Inst. of Photonics and Quantum Electronics (IPQ), Karlsruhe Institute of Technology (KIT), Germany
[4]Institute of Microstructure Technology (IMT), Karlsruhe Institute of Technology (KIT), Germany

[5]Dipartimento di Elettronica e Informazione, Politecnico di Milano, 20133 Milano, Italy

[6]Department of Electrical, Computer and Energy Engineering, University of Colorado, Boulder, USA

[7]Univ. Paris-Sud, Institut d'Electronique Fondamentale, UMR8622, Orsay, F-91405, France
[8]CNRS, Orsay, F-91405, France

[*]To whom correspondence should be addressed: roel.baets@ugent.be

Abstract

*In the article "Nonreciprocal Light Propagation in a Silicon Photonic Circuit" (Science 333, 729-733 (2011) a nonreciprocal waveguide system based on a combination of silicon, dielectrics and metals is reported. In this Comment it is explained that the interpretation with respect to nonreciprocity in this paper is incorrect and conflicts with the fundamental Lorentz reciprocity theorem. It is further pointed out that a previous publication already introduced the device concept.*

In the article "Nonreciprocal Light Propagation in a Silicon Photonic Circuit" by Liang Feng and co-workers [1] a silicon waveguide with a periodic perturbation consisting of dielectrics and metals is analyzed numerically, fabricated and tested. It is claimed that this device has nonreciprocal light transmission properties. This is assumed to be due to the special design of the grating which has a one-sided (asymmetric about $k = 0$) spatial frequency spectrum. The argumentation is supported by plots of simulated and experimental field distributions within the device. It is claimed that the nonreciprocal properties allow to implement on-chip optical isolators without the use of magneto-optical materials or nonlinear effects or time-dependent modulation.

While the numerical and experimental results seem correct, the interpretation presented in the paper is in direct conflict with the Lorentz reciprocity theorem in electromagnetism. The Lorentz reciprocity theorem [2] can be derived rigorously from Maxwell's equations. As a consequence of this theorem the frequency domain scattering matrix of a source-free linear time-independent (LTI) electromagnetic system is symmetric if all materials within the system have symmetric permittivity and permeability tensors. This property holds even if the system has no geometrical symmetry planes, and holds also for systems with loss and/or gain (complex permittivity). Critical to the understanding of the scattering matrix is that it relates outgoing scattering amplitudes at certain ports due to ingoing excitation amplitudes at other (or the same) ports, *where a port is a mode of the optical system at a given cross-section, not e.g. total power through a particular cross-section.* The structure studied in the article consists of silicon, silicondioxide, germanium and chromium. All these materials are known to have symmetric permittivity tensors (and diagonal permeability tensors), and indeed in the article are treated as scalars. Hence the interpretation is in conflict with the reciprocity theorem. In LTI-systems reciprocity can only be broken by using non-reciprocal materials, i.e. materials with a non-symmetric permittivity or permeability tensor. The only known materials with this property are magnetooptical materials in the presence of a DC magnetic bias field, such as for example yttrium iron garnet (YIG). The magnetic field breaks the time-reversal symmetry in such materials as a result of the Lorentz-force (not surprisingly the same Hendrik Lorentz who derived the reciprocity theorem). In nonlinear systems or in time-dependent (i.e. modulated) systems it is well known that one can achieve nonreciprocal behavior in absence of magnetic materials or magnetic field [3,4]. However, the system in [1] is an LTI-system.

The waveguide system proposed in the article consists of a bimodal waveguide perturbed by a grating, as shown in fig.1a. The waveguide supports two guided modes hereafter called the A-mode (being the fundamental mode) and the B-mode. If the waveguide is transversely symmetric these two modes are symmetric and anti-symmetric respectively. Such a structure can be considered to be a four-port as shown in fig.1b.

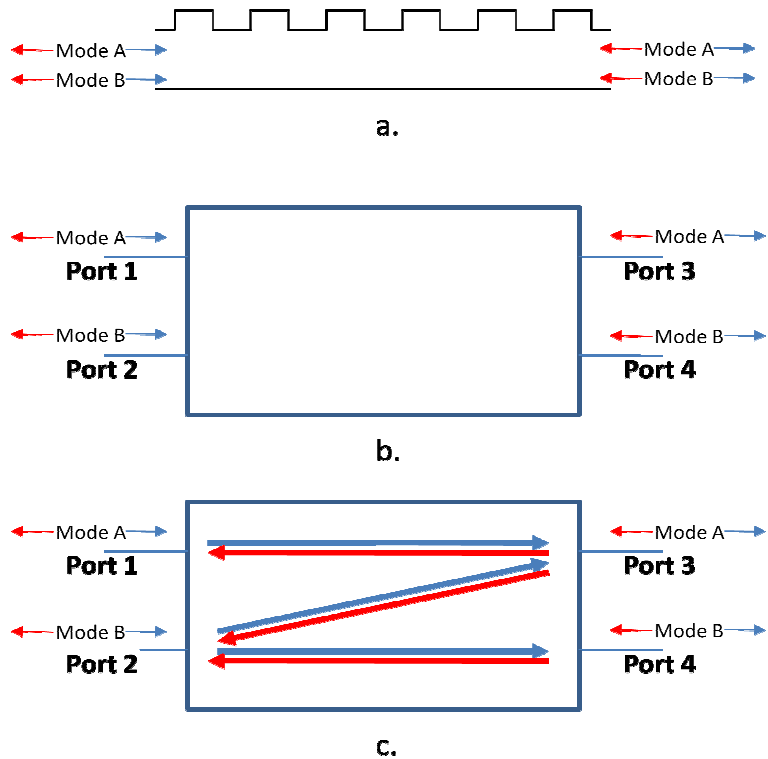

*Figure 1 Schematic diagram of a two-moded waveguide with grating (a), its general black-box description (b), amplitude transfer mechanisms in a single-sideband grating assisted codirectional coupler (c). Power transfer from left to right is indicated in blue while power transfer from right to left is indicated in red.*

At the left side the two ports represent A- and B-modes propagating into or out of the left side of the waveguide and likewise for the right side. Since the structure is linear, time-independent and source-free, it can be described fully with respect to the four ports of interest by a 4x4 scattering matrix S as a function of optical frequency f:

$$S(f) = \begin{bmatrix} s_{11}(f) & \cdots & s_{14}(f) \\ \vdots & \ddots & \vdots \\ s_{41}(f) & \cdots & s_{44}(f) \end{bmatrix}$$

The grating has a period Λ such that the two forward modes can couple with each other through phase matching and likewise the two backward modes can couple. Such a structure is known as a codirectional grating-assisted coupler. If the grating is implemented as a periodic perturbation of lossless dielectric materials – i.e. if it is an index grating – one can write the refractive index in a section through the grating as:

$$n(z) = n + \Delta n \cos(qz) = n + \frac{\Delta n}{2}[\exp(jqz) + \exp(-jqz)]$$

whereby the spatial frequency q equals 2π/Λ. Harmonics of the base frequency have been left out for simplicity here. This equation indicates that the grating has a spatial Fourier spectrum with both a positive and a negative frequency component with equal strength (a symmetric spectrum).. Due to this perturbation there is a *mutual* (and equally strong) coupling between the two forward modes. The same holds for the two backwards modes. This is schematically depicted in fig.2a, in which the propagating waves are represented by their k-vectors and the grating by the q-vector.

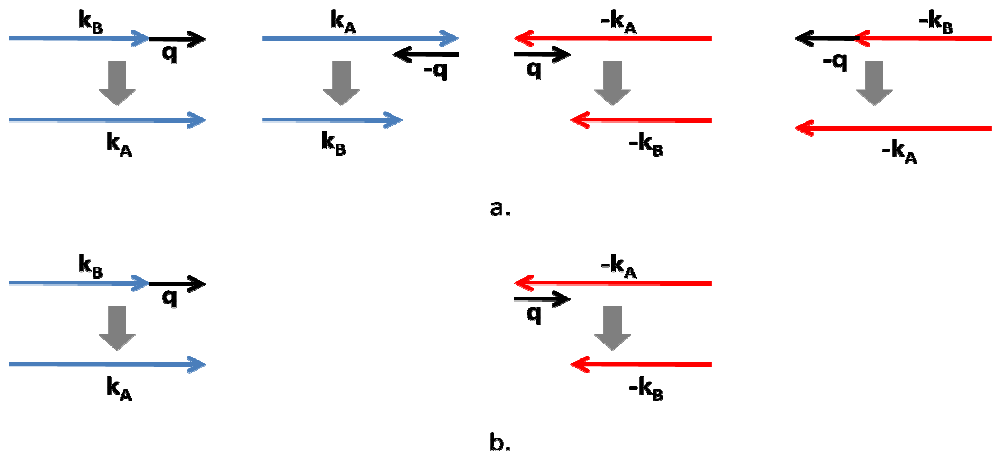

*Figure 2. Wave coupling mechanisms for an index grating (a) and for a single-sideband grating (b).*

In the paper a special grating is considered, which is a combination of an index grating $\Delta n \cos(qz)$ and a gain/loss grating $j \Delta n \sin(qz)$ such that:

$$n(z) = n + \Delta n \exp(jqz)$$

It is a "single-sideband" grating and it only allows the two coupling mechanisms in fig.2b (*or their complementary counterparts depending on the sign of q*). The two coupling mechanisms shown in fig.2b represent each other's reciprocal counterpart. The working principle of such a grating has been already published by Greenberg and Orenstein in 2004-2005 [5-7], where it is called "irreversible and unidirectional power transfer in a waveguide system". It is important to emphasize that this structure has no symmetry plane, since the function exp(jqz) does not have a mirror point. Hence it is not surprising that the structure has a completely different behavior for light injected from the left as compared to light injected from the right. The A-mode injected from the right may couple to the B-mode at the left output, but the A-mode injected from the left will not couple to the B-mode at the right. This is exactly the situation described in the article and explains the numerical and experimental field plots.

However this has *nothing* to do with nonreciprocity, and further does not provide a tool for building an isolator (unless one wishes to call a reciprocal attenuator also an isolator). If reflections are neglected the S-matrix for this structure will look like:

$$S = \begin{bmatrix} 0 & 0 & s_{13} & 0 \\ 0 & 0 & s_{23} & s_{24} \\ s_{31} & s_{32} & 0 & 0 \\ 0 & s_{42} & 0 & 0 \end{bmatrix}$$

whereby $s_{13}= s_{31}$, $s_{24}= s_{42}$ and $s_{23}= s_{32}$ in structures with reciprocal materials, as implied by the Lorentz reciprocity theorem.

This S-matrix is schematically shown in fig.1c. The scheme clearly indicates that the propagation occurring in the structure for the A-mode being injected from the left is very different from the situation where the A-mode is injected from the right. But the fact remains that $s_{13}= s_{31}$ and hence the transmission of the fundamental mode is reciprocal. This means that when the A-mode comes in from the left and is partially transmitted as the A-mode leaving the right, should this mode encounter a reflection outside the circuit and re-enter the device from the right as A-mode it will necessarily come back to the left as an A-mode with the same transmission efficiency, thus exhibiting reciprocal behavior, not isolator behavior (unless one considers simple attenuation also to be a form of isolation). It is somewhat intriguing to figure out how power conservation works in both cases. Referring again to the case of fig.1c, when mode A comes in at port 3 its power will be transmitted to port 1 and 2 and some fraction may be absorbed because the grating contains absorbing materials. When mode A comes in at port 1 the same fraction of light will remain in the fundamental mode and leave port 3 as in the first case, but no light will couple to the B-mode. Hence a bigger fraction of light will be absorbed than in the first case. This is not surprising since the field distribution in both cases is completely different and hence the field overlap with the periodic teeth of absorbing material is completely different.

It is useful to mention that when an A-mode is injected from the left and leaves the circuit at the right side as an A-mode and when subsequently this mode encounters a very special reflection outside the circuit that converts it into a B-mode reentering the circuit at port 4, then no light will leave the circuit as A-mode at port 1. So this very special case provides perfect isolation. However, an isolator has to block all reflecting light, irrespective of the nature of the reflection. The addition of mode filters or polarizers can bring no solution here.

In order to claim non-reciprocity in this device the authors should quantify numerically or experimentally whether the right-to-left transfer coefficient of the fundamental A-mode differs from the left-to-right one. The shown near field images in the article do not prove anything of the kind. A correct decomposition of the field in the right-to-left direction will reveal that the fraction of light remaining in the fundamental mode is equal to the fraction of fundamental mode being transferred in the left-to-right direction.

As a conclusion one can state that a linear and time-independent silicon waveguide system, consisting of non-magnetic dielectric materials and metals, is always reciprocal and cannot be used for isolator or circulator functionality (any more than an attenuator can be used as an isolator). Hence the claims in the article by Feng et al [1] are incorrect and the device they propose is no more an isolator than any simple attenuator. We note further that the working principle of the device described in [1] has been published by Greenberg and Orenstein in 2004.